\documentstyle[12pt]{article}
\begin{document}

\begin{flushright}
{\sf Portsmouth University \\
Relativity and Cosmology Group \\
{\em Preprint} RCG 95/13}
\end{flushright}
\[ \]
\begin{center}
{\Large \bf Inhomogeneous universes in observational coordinates}
\[ \]
Roy Maartens${}^{\dag}$\footnote{Member
of Centre for Nonlinear Studies, Witwatersrand
University, South Africa},
Neil P Humphreys${}^{\dag}$,
David R Matravers${}^{\dag}$
and William R Stoeger${}^{\ddag}$\\
\[ \]
\end{center}
${}^{\dag}${\footnotesize School of Mathematical Studies, Portsmouth
University, Portsmouth PO1 2EG, England}\\
${}^{\ddag}${\footnotesize Vatican Observatory Research Group,
Steward Observatory, Arizona University, Tucson AZ 85721, USA}
\[ \]
PACS numbers: 98.80.Hw, 98.80.Es, 04.20.Jb
\[ \]
\begin{center}
{\bf Abstract}\\
\end{center}

\noindent Isotropic inhomogeneous dust universes
are analysed via observational coordinates
based on the past light cones of the observer's galactic worldline.
The field equations are reduced to a single first--order {\sc ode}
in observational variables on the past light cone,
completing the observational integration
scheme. This leads naturally to an explicit
exact solution which is locally nearly homogeneous
(i.e. {\sc frw}), but
at larger redshift develops inhomogeneity. New observational
characterisations of homogeneity ({\sc frw} universes) are also given.

\newpage
\section*{1. Introduction}

The high degree of isotropy of the cosmic
microwave background radiation is usually taken as evidence that
the late (matter--dominated) universe is isotropic about our
worldline. When combined with the Copernican principle that we
do not occupy a privileged position, this leads to isotropy about
all worldlines, and thus to a Friedmann--Robertson--Walker ({\sc frw})
universe. (The proof of this background--radiation
argument is the Ehlers--Geren--Sachs
theorem \cite{egs}, whose perturbed version is proved in \cite{sme}.)

If we suspend the Copernican
assumption in favour of a directly observational approach, as set
out in detail in \cite{enm}, then it turns out that isotropy of
the background radiation is insufficient to force isotropy onto
the spacetime geometry. However, it has been shown \cite{mm} that
isotropy of galactic number counts and
area distances
(equivalently luminosity distances), together with the vanishing
of cosmological proper motions and image distortion,
does force isotropy of the
universe about our worldline. The universe is then a
Lema\^{\i}tre--Tolman--Bondi ({\sc ltb})
model \cite{wb}, including the special case of homogeneous {\sc frw}
models.

Current galactic observations are nowhere near the accuracy and
completeness of the background radiation observations, but they are
not inconsistent with isotropy. However, observations of the
galactic distribution do not imply homogeneity \cite{e},\cite{p};
indeed, they sometimes
appear to indicate that inhomogeneities may not be
smoothed out on larger and larger averaging scales. Therefore,
motivated by observations, and leaving aside the Copernican
principle, we assume isotropic galactic
observations which are not necessarily (i.e. without further
observational evidence) homogeneous. Observational
characterisations of homogeneity are then needed in order to
test the basis on which the standard {\sc frw} cosmologies rests. In
\cite{enm} it was shown that an {\sc ltb} universe is
homogeneous if and only if the observed galactic area distances
and number counts take precisely the {\sc frw} form
(as functions of redshift). In Section 4, we derive a further
characterisation of homogeneity within the observational
framework, and show that for the parabolic case, the
characterisation of \cite{enm} may be weakened: the single
(area distance, number count) relation must take the
Einstein--de Sitter form.

The $3+1$ analysis of {\sc ltb} spacetimes \cite{wb},
based on spatial hypersurfaces $\{t={\rm const.}\}$,
is effective for finding and analysing exact solutions, but is
not directly suited to observational cosmology. Since observations
lead to data not on $\{t={\rm const.}\}$, but
on the past light cone of the observer, a $2+2$
approach is directly adapted to the analysis of {\sc ltb}
universes from the viewpoint of observational cosmology.
Furthermore, it is not possible simply to transform the
$3+1$ {\sc ltb}
solutions to observational coordinates, since such a transformation
requires solution of the null geodesic equation,
which is in general
only possible numerically. In observational coordinates, the
null geodesics are known directly, but the field equations are
considerably more complicated, and the analogue of the $3+1$ exact
solutions is not known. In Section 2, we briefly review the
observational coordinate approach.

This
paper complements and corrects \cite{sen}, where it is indicated
how in principle the field equations can be integrated using
the redshifts, number counts and area distances as data. An
important technical mistake was made in \cite{sen} (see Section 4),
which severely over--restricts the solutions obtained by that
procedure. Here, in Section 3, we give an explicit
reduction of the field equations in $2+2$ observational
coordinates to a single first--order ordinary differential
equation ({\sc ode}) on the past light cone. The variables in
the {\sc ode} are
the area distance and a number count function.
This corrects the integration procedure of \cite{sen}, and is
parallel to the $3+1$ case of an explicit general solution.

An application of this reduction
is the construction in Section 5 of
a class of explicit exact solutions
representing an observationally--based
inhomogeneous generalisation of flat {\sc frw} universes.
The solutions
approach the Einstein--de Sitter solution near the observer, so
that they are locally homogeneous (for small
redshift). At greater redshifts, the
solutions deviate appreciably from the Einstein--de Sitter solution:
the matter distribution is locally homogeneous, but develops
inhomogeneity at larger redshifts.

\section*{2. Metric and observations in observational coordinates}

In $3+1$ coordinates based on spatial hypersurfaces
$\{t={\rm const.}\}$, the metric is \cite{wb}
\begin{equation}
ds^2=-dt^2+\left[{\partial R(t,r)\over\partial r}\right]^2
{dr^2\over 1-kf(r)^2}+R(t,r)^2d\Omega^2
\label{tb}
\end{equation}
where $k=0,1,-1$ corresponds, respectively, to
parabolic, elliptic and hyperbolic
intrinsic geometry of the $\{t={\rm const.}\}$ hypersurfaces,
and $d\Omega^2$ is the
metric of the unit 2--sphere. The coordinates are comoving
with the dust four--velocity $u^\mu=\delta^\mu{}_t$.
If $r$ is chosen as a proper area
coordinate, then the {\sc frw} case is given by
$$
R(t,r)=ra(t)~,~~f(r)=r
$$
In the general case, the solution $R(t,r)$ of Einstein's equations
may be given explicitly \cite{wb}. However these coordinates and
solution are not adapted to observations, which are made not on a
$\{t={\rm const.}\}$ surface, but rather on our past light cone. (See
\cite{enm},\cite{e},\cite{r} for further discussion of the crucial
implications of this point.) Light emitted from
a source at radial distance $r_E$ at cosmic time $t_E$ propagates
along the null geodesic $\sqrt{1-kf^2}dt=-(\partial R/\partial r)dr$
before being observed
at $r=0$ at time $t_O>t_E$. The explicit analytic form $t=t(r)$ of
the null geodesic cannot in general be determined.
If we use observational
coordinates based on the past light cones of the observer's
worldline, then the observations are given in the most simple
and direct form. The price paid for this is that the field
equations can no longer be solved explicitly. However, as we
shall show, they may be reduced to a single first--order {\sc ode}.

In observational coordinates $x^\mu=(w,y,\theta,\phi)$, the metric
takes the $2+2$ form \cite{sen}
\begin{equation}
ds^2=-A(w,y)^2dw^2+2A(w,y)B(w,y)dwdy+C(w,y)^2d\Omega^2
\label{oc}
\end{equation}
where $\{w={\rm const.}\}$ are the past light cones along
$\{y=0\}$, and $y$ is a comoving radial distance parameter
down the light rays (null geodesics) $\{w={\rm const.},
(\theta,\phi)={\rm const.}\}$. Regularity
on $\{y=0\}$ imposes on the metric the central limiting conditions
\cite{enm}: as $y\rightarrow 0$
\begin{equation}
\begin{array}{ll}
A(w,y)=A(w,0)+O(y) & A(w,0)\neq 0 \\
B(w,y)=B(w,0)+O(y) & B(w,0)\neq 0 \\
C(w,y)=B(w,0)y+O(y^2) & {}
\end{array}
\label{cent}
\end{equation}
The nontrivial coordinate freedom is a scaling of $w$ and $y$:
\begin{equation}
w\rightarrow \tilde{w}=\tilde{w}(w)~,~~y\rightarrow\tilde{y}=
\tilde{y}(y)~~~~({d\tilde{w}\over dw}\neq 0 \neq {d\tilde{y}\over dy})
\label{wy}
\end{equation}
The first corresponds to a freedom to choose $w$ as any time parameter
(e.g. proper time) along $\{y=0\}$. The second corresponds to a
freedom to choose $y$ as any null distance parameter (e.g. affine
parameter, redshift) on an initial light cone $\{w=w_0\}$, after
which $y$ is dragged along by the matter flow. These scalings
imply a scaling of the metric functions $A,B$:
$$
A\rightarrow\tilde{A}={dw\over d\tilde{w}}A~,~~
B\rightarrow\tilde{B}={dy\over d\tilde{y}}B
$$

The metric function $C$ is the observer area
distance (or angular diameter distance) on each past light cone,
since $dS_E=C^2d\Omega_E$, where $dS_E$ is the cross
sectional area of the source and $d\Omega_E$ the solid angle
it subtends at the central
observer $\{y=0\}$ \cite{enm}. If $d_A$ denotes the area
distance that is measured here--and--now, then
\begin{equation}
d_A(y)=C(w_0,y)
\label{da}
\end{equation}
where $y$ may be chosen as the redshift, so that observations
determine $C$ on $\{w=w_0\}$. By contrast, in the
$3+1$ approach, $d_A(r)=R(t(r),r)$, where $R$ is known from the
explicit exact solution, but $r$ is not known and neither is
the explicit form $t(r)$ of the past light cone of here--and--now.
The luminosity distance is
\cite{enm}
\begin{equation}
d_L=(1+z)^2 d_A
\label{d}
\end{equation}
where the redshift $z$ of the source relative
to the central observer is given by \cite{enm}:
\begin{equation}
1+z={A(w_0,0)\over A(w_0,y)}
\label{z}
\end{equation}

The metric function $B$ determines the deviation of $y$ from the
affine parameter $\nu$ \cite{enm}
\begin{equation}
B={1\over A}{d\nu\over dy}
\label{B}
\end{equation}
Two four--vectors are defined intrinsically
by the matter and light
rays: the dust velocity and photon wave vectors \cite{enm}
\begin{equation}
u^\mu = A^{-1}\delta^\mu{}_w ~,~~
k^\mu = (AB)^{-1}\delta^\mu{}_y
\label{uk}
\end{equation}

The number of galactic sources counted by
the central observer out to a distance $y$ is
[see \cite{sen}, which corrects the error in \cite{enm},
equation (19)]
\begin{equation}
N(y)= 4\pi\int_0^y n(w_0,\tilde{y})B(w_0,\tilde{y})C(w_0,\tilde{y})^2
d\tilde{y}
\label{N}
\end{equation}
on using (\ref{z}) and (\ref{B}). In (\ref{N}),
selection effects are ignored, the freedom (\ref{wy}) has been used
to set $A(w_0,0)=1$, and
$n$ is the number density of sources, so that the matter density
is
$$\rho=4\pi mn$$
where $4\pi m$ is the average galactic mass.

The rate of expansion of the dust is
$3H=u^\mu{}_{;\mu}$, so that, by (\ref{oc}),
\begin{equation}
H={1\over 3A}\left({\dot{B}\over B}+2{\dot{C}\over C}\right)
\label{h}
\end{equation}
where a dot denotes $\partial/\partial w$. For the central observer,
relative to whom the recession of other worldlines is
isotropic, $H$ is precisely the Hubble rate.
In the homogeneous {\sc frw}
case, $H$ is constant at each instant of time $t=t_0$.
But in the general
inhomogeneous case, $H$ varies with radial distance from $r=0$
on $\{t=t_0\}$. By (\ref{cent}), the central behaviour of $H$ is
$$
H(w,y)=\left[{1\over A(w,0)}{\dot{B}(w,0)\over B(w,0)}\right]+O(y)
$$
At any given instant $w=w_0$ along $\{y=0\}$, the first term is just
the Hubble constant $H_0=H(w_0,0)$ measured by the central observer:
\begin{equation}
H_0={1\over A_0}{\dot{B}_0\over B_0}
\label{hcent}
\end{equation}

The covariant four--velocity of the dust is given by the gradient
of proper time along the matter worldlines: $u_\mu=-t,_\mu$. It is
also given by (\ref{uk}) and (\ref{oc}) as
$$
u_\mu=g_{\mu\nu}u^\nu=-Aw,_\mu+By,_\mu
$$
Comparing these two forms, we get
\begin{equation}
dt=Adw-Bdy~~\Leftrightarrow~~A={\partial t\over\partial w}~,~~
B=-{\partial t\over\partial y}
\label{tw}
\end{equation}
which shows that the surfaces of simultaneity for the observer
are given in observational coordinates by $Adw=Bdy$. The
integrability condition of (\ref{tw}) is
\begin{equation}
A'+\dot{B}=0
\label{ab}
\end{equation}
where a prime denotes $\partial/\partial y$. This is precisely the
momentum conservation equation (see \cite{sen}).

Using (\ref{cent}), (\ref{da}), (\ref{z}), (\ref{hcent})
and (\ref{ab}), it follows that, exactly as in the {\sc frw} case, the
Hubble constant is determined by the initial
slope of the $(d_A,z)$ relation,
or equivalently of the $(d_L,z)$ relation:
\begin{equation}
H_0^{-1}=
\left({dd_A\over dz}\right)_0=
\left({dd_L\over dz}\right)_0
\label{hd}
\end{equation}

Since the acceleration and vorticity of the dust are zero, the
only non--trivial kinematic quantities are the expansion (\ref{h})
and the shear $\sigma$. Using (\ref{oc}),
(\ref{h}) and (\ref{ab})
in the general expressions given in \cite{mm}, we find
\begin{equation}
\sigma ={2\over \sqrt{3}A}\left|{\dot{B}\over B}-{\dot{C}\over C}
\right|
\label{shear}
\end{equation}
Thus the {\sc frw} case is given by $\dot{B}/B=\dot{C}/C$.

\section*{3. Reduction of the field equations}

The conservation equations are (\ref{ab})
and the energy conservation equation $u^\mu\nabla_\mu(\rho)+3H\rho=0$,
which gives, by (\ref{N}) and (\ref{h})
\begin{equation}
\rho={mN'\over BC^2}
\label{rho}
\end{equation}
The central behaviour of the density is
$\rho(w,y)=\rho(w,0)+O(y)$,
and then (\ref{rho}) and (\ref{cent}) imply
\begin{equation}
\begin{array}{l}
\rho(w,y)=\left[{\rho_0B_0^3\over B(w,0)^3}\right]+O(y) \\
{} \\
N(y)=\left[{\rho_0B_0^3\over 3m}\right]y^3+O(y^4)
\end{array}
\label{mcent}
\end{equation}

In observational coordinates, the field equations
lead to the following first integrals (see \cite{sen}):
\begin{eqnarray}
{\dot{C}\over A}+{C'\over B} &=& F\equiv {N_*'\over N'} \label{c.'}\\
{\dot{C}\over C}{C'\over C}+{A\over 2B}{C'^2\over C^2}
-{AB\over 2C^2} &=& -{mN_*\over 2C^3}  \label{g}
\end{eqnarray}
where $N_*(y)$ is an arbitrary function, whose central behaviour is
the same as that of the number counts:
\begin{equation}
N_*(y)=\left[{\rho_0B_0^3\over 3m}\right]y^3+O(y^4)
\label{gcent}
\end{equation}
by (\ref{c.'}), (\ref{cent}) and (\ref{mcent}).

In \cite{sen}, an integration scheme for the field equations, with
observational data on the past light cone $\{w=w_0\}$, is presented,
intending to give a formal exact solution. Due to the
over--restrictive nature of the condition $B=A$, it fails to do so.
Here we give a somewhat different approach, which parallels
the explicit exact solution in $3+1$ coordinates, and which
corrects \cite{sen}.
In terms of the intrinsically defined variables [see (\ref{uk})]
\begin{equation}
u\equiv u^\mu\nabla_\mu (C)={\dot{C}\over A}~,~~
v\equiv \left(u^\nu k_\nu\right)^{-1}k^\mu\nabla_\mu (C)={C'\over B}
\label{uv}
\end{equation}
(\ref{c.'}) and (\ref{g}) become
\begin{eqnarray}
u+v &=& F\nonumber\\
v^2+2uv &=& 1+{mN_*\over C} \nonumber
\end{eqnarray}
Solving and eliminating $u,v$, we obtain
\begin{eqnarray}
A &=& {\dot{C}\over
\left[F^2-1+mN_*/C\right]^{1/2}} \label{a}\\
B &=& {C'\over F-\left[F^2-1+mN_*/ C\right]^{1/2}} \label{b}
\end{eqnarray}
Thus the problem is reduced to determining $C$. Integrating
(\ref{a}) along the matter flow $y=$const, and using (\ref{tw}),
we get
\begin{equation}
t+T(y)=\int {dC\over
\left[F^2-1+mN_*/C\right]^{1/2}}
\label{bs}
\end{equation}
where $T$ is arbitrary. This is the {\sc ltb} exact solution
\cite{wb}, provided we identify
\begin{equation}
F^2=1-kf^2~,~~k=0,\pm 1
\label{f}
\end{equation}
Of course, we could have started from this known solution,
but then we would have to evaluate some of the
field equations in observational
coordinates anyway in order to determine $A$ and $B$. Furthermore,
the observational coordinates allow us to incorporate observations
directly into the field equations.

By (\ref{f}), $N_*$ is a modified number count variable:
\begin{equation}
N_*(y)= \int_0^y \sqrt{1-kf(\tilde{y})^2}
N'(\tilde{y})d\tilde{y}
\label{G}
\end{equation}
so that $k=0\Rightarrow N_*=N$.

Now the cosmic time $t$ in (\ref{bs}) is not known explicitly in
observational coordinates, so we must transform (\ref{bs}) into an
equation involving only observational variables. This entails losing
the explicit exact $3+1$ general solution,
but the balancing advantage, absent
in the $3+1$ case, is that the light rays are known explicitly, and
the fundamental equation involves only observational variables.

We evaluate the integral in (\ref{bs}),
differentiate down the light rays,
and use (\ref{tw}), (\ref{b}) and (\ref{G}), to get:
\begin{equation}
{C'\over 1-\sqrt{mN\over C}}+{2\over 3}\left(
\sqrt{{C^3\over mN}}\right)'-T'=0~,~~~{\rm for}~k=0
\label{one1}
\end{equation}
\begin{equation}
{C'\over\sqrt{1-kf^2}-\sqrt{{mN_*\over C}-kf^2}} =T'-
{\textstyle{1\over2}}\left\{
\begin{array}{ll}
\left[mN_*(\Gamma-\sin\Gamma)/f^3\right]' &{\rm where}\\
{}~~ \Gamma \equiv 2\arcsin(f\sqrt{C/mN_*}) & {\rm for}~k=1\\
{}&{}\\
\left[mN_*(\sinh\Gamma-\Gamma)/f^3\right]' &{\rm where}\\
{}~~\Gamma \equiv
2{\rm arcsinh}(f\sqrt{C/mN_*}) & {\rm for}~k=-1
\end{array}
\right.
\label{one}
\end{equation}
For each $k$, this is the fundamental equation,
representing the observational
reduction of the field equations, analogous to the $3+1$ case.
It is a first--order {\sc ode} for
the area distance $C$
as a function of $y$ on each light cone. Though the arbitrary
function $T'(y)$ generally leads to an infinite family of
solutions, on our past light cone $\{w=w_0\}$ this fundamental
equation relates $T'(y)$ to observationally determined
quantities. The observational data on $\{w=w_0\}$ may be taken as
$d_A(z)$ and $N(z)$.
In summary, we
can determine the arbitrary functions $f(y)$, $N_*(y)$ and
$T'(y)$ from the data in the following way, very similar to what
is suggested in \cite{sen}:\\

\noindent{\bf (a)} Set $B=A$ on our past light cone. This amounts
to a choice of $y$ on $\{w=w_0\}$,
and is valid since we do not specify
$B=A$ for all $w$. (If we do not use this freedom, then it seems
very difficult, if not impossible, to use all the available data
on our past light cone in determining the solution.)\\
{\bf (b)} Then integrate the Ehlers (or null Raychaudhuri)
equation on our past light cone, as outlined in \cite{sen}
[equations (31) to (33)]. This enables us to change our data
from functions of $z$ (the observed form) to functions of $y$
on our past light cone.\\
{\bf (c)} We can also determine $\dot{C}$ on our past light cone,
as outlined in \cite{sen} (Section 5 iii). $C'$ is essentially
given on our past light cone by the data. Thus, since we know
both $A$ and $B$ on our past light cone, we can determine
$f(y)$ from (\ref{c.'}).\\
{\bf (d)} Knowing $f(y)$, we can use (\ref{G}) to
determine $N_*(y)$ since we know $N(z)$ from data and thus
$N(y)$ from (b).
Thus we have determined
the arbitrary functions $f(y)$ and $N_*(y)$ by using the data we
presume we have available on our past light cone.\\
{\bf (e)} Finally, as indicated above, we can use (\ref{one1}),
(\ref{one}) on $\{w=w_0\}$ to find $T'(y)$. \\

\section*{4. Characterisation of homogeneity}

Observational characterisations of homogeneity (as opposed to
unverifiable philosophical assumptions such as the Copernican
principle \cite{e}), are complicated by the fact that even the
homogeneous {\sc frw} universes {\em appear} inhomogeneous - because
observations are made down the light cone, which is a hypersurface
of inhomogeneity \cite{enm},\cite{e},\cite{r}. Homogeneity of an
{\sc frw} universe is implied indirectly
by the specific forms of the observed
quantities. Any deviation from these forms indicates real
inhomogeneity (i.e. on spatial hypersurfaces).
A precise observational characterisation of homogeneity within
the {\sc ltb} class of universes has been given
\cite{enm},\cite{sen}:\\
{\em isotropic dust universes are homogeneous
if and only if the (area distance, redshift) and
(number count, redshift) relations measured by the central
observer take exactly the {\sc frw} form.}

This characterisation may be slightly weakened in the parabolic
($k=0$) case.
In the fundamental equation (\ref{one1}),
we may transform the independent variable to $N$:
\begin{equation}
\left[1-\sqrt{mN/C}\right]^{-1}{\partial C\over\partial N}+
{\partial\over\partial N}\left({\textstyle{2\over3}}\sqrt{C^3/
mN}\right)-{\partial T\over\partial N}=0
\label{cn}
\end{equation}
It is well known \cite{wb} that $T=$ const. reduces
the metric to {\sc frw}.
The first order {\sc ode} (\ref{cn}) then shows that
$d_A(N)=C(w_0,N)$ takes the {\sc frw} form
(see section 5 for the explicit form) if and only if $T=$ const.
Thus, for $k=0$, one does not need the (area distance, redshift)
and (number count, redshift) relations separately to take the
{\sc frw} form, but only the single (area distance, number count)
relation:\\
{\em a parabolic {\sc ltb} universe is Einstein--de Sitter
if and only if the (area distance, number count) relation
measured by the central observer takes
exactly the Einstein--de Sitter form.}\\
This result is consistent with \cite{enm},
but it follows more directly via our approach. Note that the result
does not extend to $k\neq 0$, essentially since in that
case 2 arbitrary functions ($T$ and $f$) need to be fixed by
2 independent observational relations.

We emphasize the `global' nature of these characterisations, i.e.
the fact that $d_A(z)$ and $N(z)$ must be {\sc frw} for {\em all}
$z$. Local observational parameters derived from $d_A$ and $N$ are
insufficient to characterise homogeneity. For example, the Hubble
constant, derived from $d_A$ as in (\ref{hd}), cannot in itself
distinguish between inhomogeneous and homogeneous universes.
The dependence of $H_0$ on
the spatial geometry (i.e. on $k$) and on the
density $\rho_0$ at the observer,
follows from (\ref{hcent}), (\ref{a}), (\ref{b}) and
(\ref{f}) as
\begin{equation}
H_0=\sqrt{{\rho_0\over 3}-k\left[{f'(0)\over B_0}\right]^2}
\label{h0}
\end{equation}
The {\sc frw} case has $f'(0)=1$ and $B_0=a_0$ (see (\ref{a=b})
and (\ref{FRW}) below), and for a given $\rho_0$ and $k$,
the {\sc frw} $H_0$ is observationally
indistinguishable from the general {\sc ltb} $H_0$.

A particular consequence of (\ref{h0}) arises when
an {\sc ltb} under--dense region is matched to an
{\sc frw} region. The
low value for $\rho_0$
implies a low value for $H_0$ if $k=0,1$.
Thus cosmological models with a {\em locally high} Hubble constant
arising
from an {\em under--dense} inhomogeneous local region within
a global {\sc frw} universe (see for example \cite{mt}),
must have $k=-1$ in the local {\sc ltb} region.
The conditions for smooth matching
at the boundary (see \cite{bc})
then require $k=-1$ in the {\sc frw} region,
i.e. an {\em open} {\sc frw} universe.

We now return to the problem of characterising homogeneity,
and derive an alternative condition (for all $k$).
As pointed out in \cite{sex}, the assertion in \cite{sen} that
the coordinate freedom (\ref{wy}) could be used to set
\begin{equation}
B(w,y)=A(w,y)
\label{a=b}
\end{equation}
is erroneous, and in fact (\ref{a=b}) implies restrictions on
the spacetime geometry. This choice made in \cite{sen}
automatically over--restricts the solutions obtained by that
procedure to be homogeneous ({\sc frw}) solutions, as we show below.
As long as $y$ is specified to be comoving, one can set $B=A$
only on one past light cone, i.e. $B(w_0,y)=A(w_0,y)$, and not
generally. Even though the $ww$ and $wy$ components of the Lie
derivative of the metric $g_{\mu\nu}$ along $u^\mu$ are equal,
the time derivatives of $A$ and $B$ are not given by the Lie
derivatives. And there is no further freedom to specify them
arbitrarily - as that freedom has already been used up by specifying
$y$ to be comoving. If, however, $y$ is {\em not} chosen comoving,
then one could set $B(w,y)=A(w,y)$ without loss of generality.

This correction
provides another characterisation of {\sc frw} universes within the
{\sc ltb} class:\\
{\em an {\sc ltb} universe is {\sc frw}
if and only if there exists a choice of radial comoving
coordinate such that
$g_{wy}=g_{ww}$ in observational coordinates}.

The proof proceeds simply and directly
from the formalism set up in sections 2 and 3.
The assumption (\ref{a=b}) allows us to integrate (\ref{ab}) and
(\ref{c.'}), to get
\begin{equation}
A(w,y)=a(\eta)~,~~C(w,y)=a(\eta)\int_0^y \sqrt{1-kf(\tilde{y})^2}
d\tilde y~,~~\eta\equiv w-y
\label{FRW}
\end{equation}
where we have used the central condition (\ref{cent}) for $C$.
Then it follows immediately from (\ref{FRW}) that the shear
(\ref{shear}) vanishes, so that the spacetime is {\sc frw}.

For completeness, we present the {\sc frw} metric and observations in
observational coordinates.
By (\ref{FRW}) we can write the metric (\ref{oc}) as
$$
ds^2=a(\eta)^2\left[-d\eta^2+dy^2+\left(\int_0^y
\sqrt{1-kf(\tilde{y})^2}
d\tilde y\right)^2d\Omega^2\right]
$$
showing that $\eta$ is conformal time. It follows that $y$ is
conformal proper radial distance $\chi$, so that \cite{s}
\begin{equation}
\int_0^\chi \sqrt{1-kf(\tilde{y})^2}d\tilde y=
{1\over \sqrt{k}}\sinh \sqrt{k} \chi
\label{FRWf}
\end{equation}

Now (\ref{a=b}), (\ref{FRW}) and (\ref{FRWf}) reduce the field
equation (see \cite{sen})
$$
{\ddot{C}\over C} = {\dot{A}\over A}{\dot{C}\over C}
-{A^{2}\over 2C^2} + {A\over B}{\dot{C}\over C}{C'\over C}
+ {A^2\over 2B^2}{C'^2\over C^2}
$$
to
$$
2a{d^2a\over d\eta^2}-\left({da\over d\eta}\right)^2-k a^2=0
$$
which has solutions
\begin{equation}
a(w-\chi)=\alpha \left\{
\begin{array}{ll}
(w-\chi)^2 & k=0 \\
1-\cos(w-\chi) & k=1 \\
\cosh(w-\chi)-1 & k=-1
\end{array}
\right.
\label{FRWsoln}
\end{equation}
where $\alpha$ is constant and the big bang occurs at $w=\chi$.

The derivation of the exact observational formulas
in {\sc frw} universes
is immediate and transparent in observational coordinates. If
$t_0$ is the age of the universe (so that $t_0=w_0$, since $\chi=0$
for the observer), then the
constant $\alpha$ follows from (\ref{hcent}) and (\ref{FRWsoln}) as
\begin{equation}
\alpha={1\over H_0}\left\{
\begin{array}{l}
2t_0^{-3}\\
\sin t_0(\cos t_0-1)^{-2} \\
\sinh t_0(\cosh t_0-1)^{-2}
\end{array}
\right.
\label{alpha}
\end{equation}
while the density parameter $\Omega_0=\rho_0/3H_0^2$ follows from
(\ref{rho}) and (\ref{mcent}):
\begin{equation}
\Omega_0=\left\{
\begin{array}{l}
1\\
2/(\cos t_0+1) \\
2/(\cosh t_0+1)
\end{array}
\right.
\label{dp}
\end{equation}
Then (\ref{FRWsoln}), (\ref{alpha}) and (\ref{dp})
lead directly to all relations between
the redshift [see (\ref{z})], observer area distance
[see (\ref{da}), (\ref{FRW})], luminosity distance [see (\ref{d})]
and number counts [see (\ref{N})]. For example, the
number count/ luminosity distance relation is
\begin{equation}
N(d_L)={3\over 4mH_0}\left\{
\begin{array}{l}
{\textstyle{32\over3}}\left[1+\left(1-\sqrt{1+2H_0d_L}\right)/
H_0d_L\right]^3\\
\Omega_0(\Omega_0-1)^{-3/2}\left[2\chi(d_L)-\sin 2\chi(d_L)\right] \\
\Omega_0(1-\Omega_0)^{-3/2}\left[\sinh 2\chi(d_L)- 2\chi(d_L)\right]
\end{array}
\right.
\label{nd}
\end{equation}
where, for $k=\pm 1$:
\begin{equation}
\chi(d_L)=\left\{
\begin{array}{l}
\arctan\left(2\sqrt{\Omega_0-1}(1+H_0d_L)\left[ H_0d_L(2-\Omega_0)
\right]^{-1}\right)+\\
{}~~~~-\arccos\left(\Omega_0H_0d_L\left[
\sqrt{\Omega_0^2H_0^2d_L^2
+(\Omega_0-1)(8H_0d_L+4)}\right]^{-1/2}\right) \\
{}\\
\ln\left[\Omega_0H_0d_L-2\sqrt{1-\Omega_0}\sqrt{1+2H_0d_L}\right]+\\
{}~~~~-\ln\left[
H_0d_L(2-\Omega_0)-2(1+H_0d_L)\sqrt{1-\Omega_0}\right]
\end{array}
\right.
\label{xid}
\end{equation}

\section*{5. Observational generalisation of the Einstein--de Sitter
universe}

The {\sc ltb} solution (\ref{bs}) allows one to write down any
number of explicit exact inhomogeneous
generalisations of {\sc frw} solutions. One simply chooses $T(y)$.
However,
if we want an {\em observationally} based generalisation, then the
{\sc ltb} solution is not appropriate, since $t$ is not an
observational variable, and deriving the observational variables
requires integration of the null geodesics.
In the observational coordinates, the
key equation (\ref{one1}) or
(\ref{one}) is far more complicated, but it has
the advantage that it is already in observational variables.
Thus any explicit solution will be a readily
interpreted {\em observational generalisation of
an {\sc frw} universe.}

We have found one such solution in the case $k=0$, i.e. an
observational inhomogeneous generalisation of an
Einstein--de Sitter universe. The equation (\ref{cn})
can be simplified by defining the observational variable
\begin{equation}
D=\sqrt{{C\over mN}}
\label{D}
\end{equation}
Then (\ref{cn}) becomes
\begin{equation}
6N {\partial D\over \partial N}+2D+1+\left[{3\over m}
{\partial T\over\partial N}\right]\left({1-D\over D^3}\right)=0
\label{onet}
\end{equation}
By (\ref{D}), (\ref{a}), (\ref{b}) and (\ref{rho}), we get
\begin{equation}
\begin{array}{l}
A(w,N)=2mND^2\dot{D} \\
B(w,N)=mMD^2(D+2ND_N)/(D-1) \\
C(w,N)=mND^2 \\
\rho(w,N)=(D-1)/[m^2N^2D^6(D+2ND_N)]
\end{array}
\label{dabc}
\end{equation}

Now it is clear that the
{\sc ode} (\ref{onet}) is separable for $\partial T/\partial N=
{1\over 3}m\beta$, $\beta=$ const.
If $\beta=0$, then $T=$ const., which is the {\sc frw} case.
In the observational variables, the
Einstein--de Sitter solution following from (\ref{onet}) and
(\ref{dabc}) is
\begin{equation}
\begin{array}{ll}
A={\textstyle{1\over4}}m\dot{P}(P-N^{1/3})^2 &
B={\textstyle{1\over12}}mMN^{-2/3}(P-N^{1/3})^2 \\
{}&{}\\
C={\textstyle{1\over4}}mN^{1/3}(P-N^{1/3})^2 &
\rho=192m^{-2}(P-N^{1/3})^{-6}
\end{array}
\label{eds}
\end{equation}
where $P(w)$ is an arbitrary function, subject to
$P(w_0)=2(mH_0)^{-1/3}$,
which ensures that $\rho_0=3H_0^2$. The (area distance, number
count) relation is $d_A(N)=C(w_0,N)$. Note that by (\ref{D}),
(\ref{eds}) and (\ref{rho}), the observational variable $D$
on $\{w=w_0\}$
determines the variation in number counts in terms of the
variation in density:
\begin{equation}
{N'\over N}=D{\rho'\over\rho}
\nonumber
\end{equation}

The inhomogeneous
case $T={1\over 3}m\beta N$, $\beta\neq 0$, leads to the solution
\begin{equation}
N=Q(w) \exp\left[-\int^{\sqrt{C/mN}}
{6D^3dD\over 2D^4+D^3-\beta D+\beta}\right]
\label{geds}
\end{equation}
where $Q$ is an arbitrary function. For small number counts
[equivalently, by (\ref{cent}), (\ref{gcent}) and (\ref{G}),
for large $D$], the solution (\ref{geds}) on $\{w=w_0\}$ has the form
\begin{eqnarray}
d_A(N)&=&(mH_0^{-2})^{1/3}N^{1/3}-(m^2H_0^{-1})N^{2/3}+
{\textstyle{1\over4}}mN+\nonumber\\
{}&{}&+{\textstyle{1\over3}}\beta(m^4H_0)^{1/3}N^{4/3}
-{\textstyle{1\over6}}\beta(m^5H_0^2)N^{5/3}+O(N^2)
\label{exp}
\end{eqnarray}
where $Q(w_0)$ has been chosen so that the leading term
in (\ref{exp}) agrees with that in (\ref{eds}). The
first three terms in (\ref{exp}) are exactly the Einstein--de Sitter
form given by (\ref{eds}).
Thus by the characterisation proved in section 4, the new solution
approaches the Einstein--de Sitter solution
for small number counts and then deviates from
it for larger number counts. Choosing $y=z$ on $\{w=w_0\}$, it follows
from (\ref{rho}) and (\ref{mcent}) that
$$
N^{1/3}=B_0\left({\rho_0\over 3m}\right)^{1/3}z+O(z^2)
$$
so that small number counts are equivalent to small redshift.
This solution represents a universe that is
locally homogeneous, but for larger redshift develops real
inhomogeneity. If the Copernican principle is not invoked,
and if large--scale observations of the galactic distribution
are taken at face value, then such a model of the late universe
is not automatically ruled out.

For completeness, we give the explicit form of the integral
in (\ref{geds}), which
depends on the roots $\alpha_i$
of the quartic in the integrand \cite{gr}: \\

\noindent{\bf (a)} $\beta=\beta_\pm \equiv 13\pm{15\over2}\sqrt{3}~
\Rightarrow ~\alpha_1=\alpha_2\equiv r=\bar{r},
\alpha_3=\bar{\alpha}_4\equiv \lambda\neq \bar{\lambda}\neq r$~:

\begin{equation}
N=Q(w)
\left(1-{1\over r}\sqrt{{C\over mN}}\right)^q
\left|\left(1-{1\over \lambda}\sqrt{{C\over mN}}\right)^\xi\right|^2
\exp\left[p\left(\sqrt{{C\over mN}}-r\right)^{-1}\right]
\label{s1}
\end{equation}
where
$$
q={3r^2[2r(\lambda+\bar{\lambda})-3\lambda\bar{\lambda}-r^2]\over
(\lambda-r)^2(\bar{\lambda}-r)^2}
$$
$$
p={3r^3[-r(\lambda+\bar{\lambda})+\lambda\bar{\lambda}+r^2]\over
(\lambda-r)^2(\bar{\lambda}-r)^2}
$$
$$
\xi={-3\lambda^3\over (\lambda-r)^2(\lambda-\bar{\lambda})^2}
$$
\[ \]
{\bf (b)} $0\neq\beta\neq\beta_\pm~\Rightarrow~{\rm no}~
\alpha_i~{\rm equal}~$:\\

\begin{eqnarray}
N&=&Q(w)
\left(1-{1\over \alpha_1}\sqrt{{C\over mN}}\right)^{\beta_1}
\left(1-{1\over \alpha_2}\sqrt{{C\over mN}}\right)^{\beta_2}
\times \nonumber\\
{}&{}&\times\left(1-{1\over \alpha_3}\sqrt{{C\over mN}}
\right)^{\beta_3}
\left(1-{1\over \alpha_4}\sqrt{{C\over mN}}\right)^{\beta_4}
\label{s2}
\end{eqnarray}
where
$$
\beta_i={-3\alpha_i^3\over (\alpha_i-\alpha_j)(\alpha_i-
\alpha_k)(\alpha_i-\alpha_l)}~,~~j<k<l
$$
and $i$ is not equal to $j,k$ or $l$. \\

Further calculations show that
the qualitative behaviour of the solutions (\ref{s1}) and (\ref{s2})
falls into two types, according to whether or not
the past light cone reconverges.

\section*{6. Concluding remarks}

Motivated by the direct observational evidence for isotropy, and by
the need to identify indirect observational evidence for
homogeneity, we have completed an exact observational analysis of
isotropic dust cosmologies, i.e. {\sc ltb} spacetimes. Such an
analysis, based on the past light cone (which is the hypersurface on
which observational data is given), was begun in \cite{enm} and
continued in \cite{sen}. An important error in \cite{sen} has been
corrected here, allowing us to properly complete the integration
scheme, as described in Section 3. The main result is the reduction
of the field equations to a single {\sc ode} on the past light cone
[equations (\ref{one1}), (\ref{one})], which then allows us to show
in detail how the observations, i.e. (area distance, redshift) and
(number count, redshift) relations, provide the data needed
on the past light cone to integrate the field equations and determine
the metric and matter distribution.

The observational {\sc ode} (\ref{one1}) leads to a simplified
characterisation of Einstein--de Sitter universes via the single
(area distance, number count) relation (Section 4). Furthermore, we
give in Section 4 a new observational characterisation of homogeneity,
i.e. the condition that $B=A$ in observational coordinates. Finally,
in Section 5 we used (\ref{one1}) to construct an exact observational
generalisation of the Einstein--de Sitter universe, given by
(\ref{geds}), (\ref{exp}), (\ref{s1}) and (\ref{s2}). This spacetime
is locally homogeneous but inhomogeneous on large scales. We are not
proposing it as a realistic model of the universe, which is more
generally believed to be locally inhomogeneous but homogeneous
on large enough scales, but it is not ruled out by current galactic
observations, and the method of deriving it could be extended to
construct models with different behaviour of the inhomogeneity.

%\newpage

\end{document}